\newcount\style \style=2
\ifodd\style
\documentstyle[aas2pp4]{article}
\else
\documentstyle[12pt,aasms4]{article}
\fi 

\begin{document}

\setlength{\baselineskip}{12pt}

\newcommand\bb[1] {   \mbox{\boldmath{$#1$}}  }
\newcommand\del{\bb{\nabla}}
\newcommand\bcdot{\bb{\cdot}}
\newcommand\btimes{\bb{\times}}
\newcommand\vv{\bb{v}}
\newcommand\B{\bb{B}}
\newcommand\BV{Brunt-V\"ais\"al\"a\ }
\newcommand\iw{ i \omega }
\newcommand\kva{ \bb{k\cdot v_A}  }
\newcommand\beq{\begin{equation}}
\newcommand\eeq{\end{equation}}

\def\lta{\mathrel{\rlap{\lower 3pt\hbox{$\mathchar"218$}}
    \raise 2.0pt\hbox{$<$}}}

\def\dd{\partial}
\lefthead{Balbus \& Hawley}
\righthead{Angular Momentum Transport in Nonradiative Accretion}
\slugcomment{Submitted to ApJ.}

\title{On the Nature of Angular Momentum Transport in Nonradiative
Accretion Flows}

\author{Steven A. Balbus  and John F. Hawley}

\affil{Dept. of Astronomy, University of Virginia, PO Box
3818,\\ Charlottesville VA 22903, USA. sb@virginia.edu;
jh8h@virginia.edu}

\begin{abstract}

The principles underlying a proposed class of black hole accretion
models are examined.  The flows are generally referred to as
``convection-dominated,'' and are characterized by inward transport of
angular momentum by thermal convection and outward viscous transport,
vanishing mass accretion, and vanishing local energy dissipation.  In
this paper we examine the viability of these ideas by explicitly
calculating the leading order angular momentum transport of
axisymmetric modes in magnetized, differentially rotating, stratified
flows.  The modes are destabilized by the generalized
magnetorotational instability, including the effects of angular
velocity and entropy gradients.  It is explicitly shown that modes that
would be stable in the absence of a destabilizing entropy gradient
transport angular momentum outwards.  There are no inward transporting
modes at all unless the magnitude of the (imaginary) \BV frequency is
comparable to the epicyclic frequency, a condition requiring
substantial levels of dissipation.  When inward transporting modes do
exist, they appear at long wavelengths, unencumbered by magnetic
tension.  Moreover, very general thermodynamic principles prohibit the
complete recovery of irreversible dissipative energy losses, a central
feature of convection-dominated models.  Dissipationless flow is
incompatible with the increasing inward entropy gradient needed for the
existence of inward transporting modes.  Indeed, under steady
conditions, dissipation of the free energy of differential rotation
inevitably requires outward angular momentum transport.  Our results
are in good agreement with global MHD simulations, which find
significant levels of outward transport and energy dissipation, whether
or not destabilizing entropy gradients are present.

\end{abstract}


\section{Introduction}

Originally developed to be powerful luminosity sources, black hole
accretion models are now confronted by an embarrassing plethora of
underluminous X-ray emitters, the best known of which is the Galactic
center source Sgr A* (Melia \& Falcke 2001).  These low luminosity
objects are thought to be prime candidates for a class of theoretical
accretion models that has been intensively studied in recent years, 
which we shall refer to generically as nonradiative accretion flows.
Accretion generally requires significant energy loss, and the absence
of radiative losses in these flows means that determining the ultimate
fate of the gas is less than straightforward.

Much of the recent interest in nonradiative flows was sparked by the
work of Narayan \& Yi (1994), who examined a series of one-dimensional,
self-similar, steady accretion models referred to as
``advection-dominated accretion flows,'' or ADAFs for short.  In these
models an $\alpha$ viscosity allows angular momentum transport and the
resulting flows are quasi-spherical and substantially sub-Keplerian.
Dissipative heating increases toward the flow center, creating an
inwardly increasing entropy profile which we shall refer to as
``adverse.''

Nonradiative accretion flows are amenable to numerical simulation.
Hydrodynamical simulations carried out with large assumed $\alpha$
values ($>0.3$) show some similarities to ADAFs (\cite{ia99}, 2000),
but smaller values of $\alpha$ led to flows with substantial
turbulence, smaller than anticipated net inward mass accretion rates,
and density distributions that are far less centrally peaked than in
ADAFs (\cite{spb99}; \cite{ia99}, 2000).  Both inward and outward mass
fluxes were observed at different times and different locations within
the flow, and they nearly cancelled.

These findings were given the following interpretation by Narayan,
Igumenschev, \& Abramowicz (2000; hereafter NIA), Quataert \& Gruzinov
(2000; hereafter QG), and Abramowicz et al.~(2002; hereafter AIQN).
The adverse entropy gradient triggers an instability, and significant
levels of convection result, hence the global solutions were called
``convection-dominated accretion flows''  (CDAFs).  The next step in
the argument is key:  invoking the findings of other hydrodynamical
simulations (Stone \& Balbus 1996; Igumenschev, Abramowicz, \& Narayan
2000), the angular momentum transport generated by the convective
turbulence was claimed to be {\em inward}.  This inward transport by
convection was envisioned to be sufficiently great as to cancel the
primary outward angular momentum transport by whatever process the
$\alpha$ viscosity was modeling---presumably the magnetorotational
instability, or MRI (Balbus \& Hawley 1991).

In the CDAF scenario the vanishing of the angular momentum flux implies
that the $R\phi$ component of the stress tensor responsible for
accretion also vanishes (NIA, AIQN).  This has the further consequence
that there is essentially no mass accretion and no dissipation, despite
the presence of vigorous turbulence throughout the bulk of the flow.
The only region where there is any mass accretion in the model is at
the very inner edge of the flow.  All of the energy release associated
with this small net accretion is transported outward to infinity by the
surrounding convective flow which is maintained with no further
dissipative losses.  For this reason, CDAFs are put forth as natural
candidates to explain under-luminous X-ray sources.  These models have
been elaborated upon, becoming influential and widely-cited.  Since
black hole accretion models are central to our understanding of much of
X-ray astronomy, the theoretical foundations for CDAFs deserve careful
scrutiny.

In this work we carry out an explicit analysis of magnetized,
rotationally-supported gas in the presence of an adverse entropy
gradient.  We find that CDAF models have two major inconsistencies.
First, locally unstable disturbances with adverse entropy gradients do
not generally transport angular momentum inwards in magnetized fluids.
Rather, they generally transport angular momentum outwards.
Qualitatively, their  behavior is indistinguishable from standard MRI
modes.  This latter point has been emphasized elsewhere (Hawley,
Balbus, \& Stone 2001; Balbus 2001), but here we demonstrate it
quantitatively by explicitly calculating the leading order angular
momentum transport associated with unstable WKB modes.  Convective
modes transport angular momentum outwards when magnetic tension is
significant, and inwards only for the very longest wavelength (global
scale) disturbances, where magnetic tension forces are negligible.
Indeed, for a given wavenumber, the direction of angular momentum
transport is less a matter of whether it is destabilized by convection
or rotation, and more a matter of the nature of background medium:  is
it effectively magnetized or not?  This is the crucial issue.

The second difficulty is more direct and fundamental, affecting
magnetic and nonmagnetic models alike.  By relying upon dissipated heat
energy to trigger a convective instability that supposedly renders the
flow dissipation-free, CDAFs run afoul of thermodynamic principles.  If
the source of the free energy is differential rotation, the direction
of angular momentum transport {\em must} be outward.  This is a serious
inconsistency.  The dissipation is quite significant, and is indeed
essential if convectively unstable entropy profiles are to be
sustained.

We are led to a much more standard picture of the dynamics of turbulent
accretion flows, though one at odds with the tenets of CDAF theory.
The turbulent stress tensor in magnetized differentially rotating gas
does {\em not} vanish.  There is vigorous local turbulent dissipation.
There is mass accretion.  The near cancellation of instantaneous inward
and outward mass fluxes is a property of any turbulent flow with large
rms fluctuations, and not a superposition of the contributions from two
distinct sources of mass flux with opposite signs.

In the following sections, we present (\S 2) the details of the angular
momentum calculation showing outward transport; (\S 3) an explanation
of important thermodynamic inconsistencies evident in CDAF theory; (\S
4) a brief review of numerical simulations and a concluding
summary.

\section {Radial Angular Momentum Transport}

\subsection {Local WKB Modes} Consider a disk with radially decreasing
outward entropy and pressure gradients.  We use standard cylindrical
coordinates $(R, \phi, Z)$.  The square of the \BV frequency ($N^2$) is
thus negative, and tends to destabilize.  In what follows, it is
convenient to work with the positive quantity
\beq
{\cal N}^2 \equiv -N^2  = {3\over 5\rho}{\dd P\over\dd R}{
\dd\ln P\rho^{-5/3}\over\dd R}> 0.
\eeq

The disk is differentially rotating with decreasing outward
angular velocity $\Omega(R)$, and epicyclic frequency
\beq
\kappa^2 = 4\Omega^2 + {d\Omega^2\over d\ln R} = {1\over R^3}
{dR^4\Omega^2\over dR} > 0.
\eeq
A vertical magnetic field $\bb{B} = B \bb{e_Z}$ threads the disk.
Its associated Alfv\'en velocity is  $v_A^2 = {B^2/4\pi\rho}$,
where $\rho$ is the gas density.
Axisymmetric WKB
plane wave displacements of the form
\beq
\bb{\xi} (R, Z, t) = \bb{\xi} \exp (i{kZ} - i \omega t),
\eeq
where ${k}$ and $\omega$ are respectively the vertical wavenumber
vector and the angular frequency,
lead to the dispersion relation (Balbus \& Hawley 1991)
\beq\label{disp}
{\tilde\omega}^4 +{\tilde\omega}^2({\cal N}^2 -\kappa^2)
-4\Omega^2 (kv_A)^2 = 0 ,
\eeq
where
\beq
{\tilde\omega}^2 =\omega^2 -(kv_A)^2.
\eeq

Let $\gamma= - i\omega$.
Then, the unstable branch of the dispersion relation (\ref{disp}) is
\beq\label{growth}
\gamma^2 = -(kv_A)^2 +{1\over2} \left[ {\cal N}^2 -\kappa^2 +
\sqrt{({\cal N}^2 - \kappa^2)^2 +16\Omega^2(kv_A)^2}\right].
\eeq
It is straightforward to show that these unstable modes must have
\beq
(kv_A)^2 < {\cal N}^2 - {d\Omega^2\over d\ln R} =
{\cal N}^2 + \left|{d\Omega^2\over d\ln R}\right| ,
\eeq
and that the maximum growth rate is
\beq
\gamma_{max} = {\Omega\over 4} \left( {{\cal N}^2\over\Omega^2 } +
\left|{d\ln\Omega^2\over d\ln R}\right| \right),
\eeq
which is attained for wavenumbers satisfying
\beq\label{mg}
(kv_A)_{max}^2 = 
\Omega^2 \left( 1 - {( {\cal N}^2 -\kappa^2)^2\over 16\Omega^4} \right).
\eeq

\subsection{Stress Calculation}

The angular momentum flux is directly related to the
$R\phi$ component of the stress tensor
\beq
T_{R\phi}  = \rho ( \delta v_R\, \delta v_\phi - \delta v_{AR}\,
\delta v_{A\phi}),
\eeq
where $\delta$ denotes an Eulerian perturbation, and 
\beq
\bb{\delta v_A} = { \bb{\delta B}\over \sqrt{4\pi\rho}}.
\eeq
For the local WKB modes we consider here, the angular momentum flux is
$R\Omega T_{R\phi}$.  Hence, the sign of the transport is simply the
sign of $T_{R\phi}$.

The needed expressions can be written down immediately from the
equations (2.3c--g) of Balbus \& Hawley (1991).  In terms
of $\gamma$, they are
\beq\label{Rey}
\delta v_\phi \delta v_R = (\delta v_R^2) {\Omega\over D \gamma}
\left( 
{(kv_A)^2\over\gamma^2} \left|{d\ln \Omega\over d\ln R}\right|
- {\kappa^2\over 2\Omega^2} \right),
\eeq
\beq\label{Max}
-\delta v_{A\phi} \delta v_{AR} = (\delta v_{AR})^2 {2\Omega\over D \gamma} 
= (\delta v_{R})^2 \left(kv_A\over \gamma\right)^2  {2\Omega\over D \gamma} ,
\eeq
where 
\beq
D= 1 + {(kv_A)^2\over \gamma^2}.
\eeq
These equations are general beyond our simple example, holding in the
presence of both vertical and radial entropy gradients.  Note that
there is no explicit dependence upon ${\cal N}$; the only dependence
upon ${\cal N}$ anywhere is through the growth rate $\gamma$.  The
Maxwell stress (\ref{Max}) must always be positive.  For a given growth
rate, angular momentum transport is completely determined by rotation
and magnetic tension.

Figure (1) shows the stability and angular momentum transport
properties of an unmagnetized Keplerian disk in the $k^2-{\cal N}^2$
plane.  The vertical scale is immaterial: when ${\cal N}>\kappa$,
convectively unstable modes are triggered at all wavenumbers.  There is
nothing special about the physics of convection {\em per se} with
regard to inward angular momentum transport.  In the absence of a
magnetic field, any axisymmetric disturbance governed by the above
equations would transport angular momentum inwards.  This is why
hydrodynamic simulations consistently find inward transport.

\begin{figure}
\epsscale {1.05}
\plotone{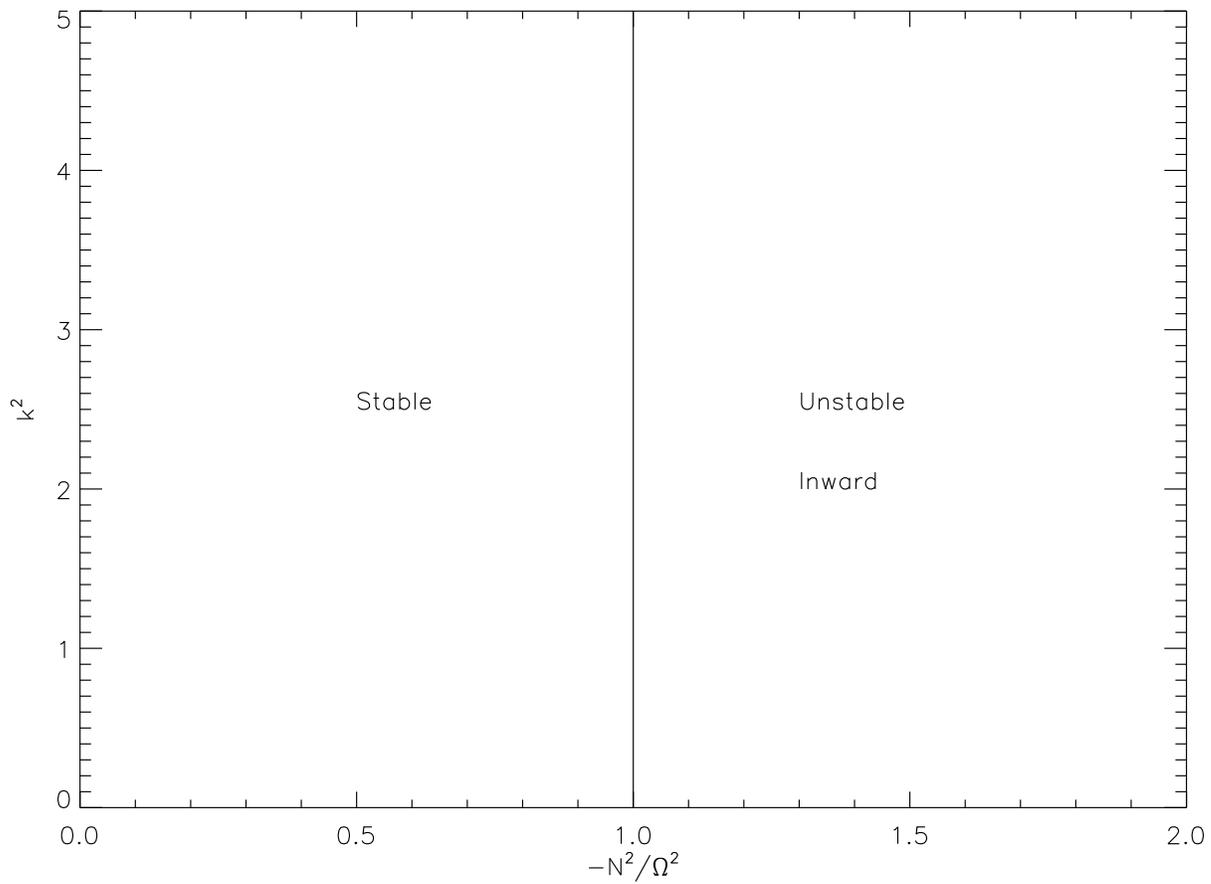}
\caption{Region of instability in the $k^2$ --  $|N^2|/\Omega^2$ plane,
for an unmagnetized disk with a Keplerian rotation profile.  The system
is unstable for all wavenumbers when $|N^2| > \kappa^2$, and the sense
of transport is inward.}\label{stability} \end{figure}

\begin{figure}
\epsscale {1.05}
\plotone{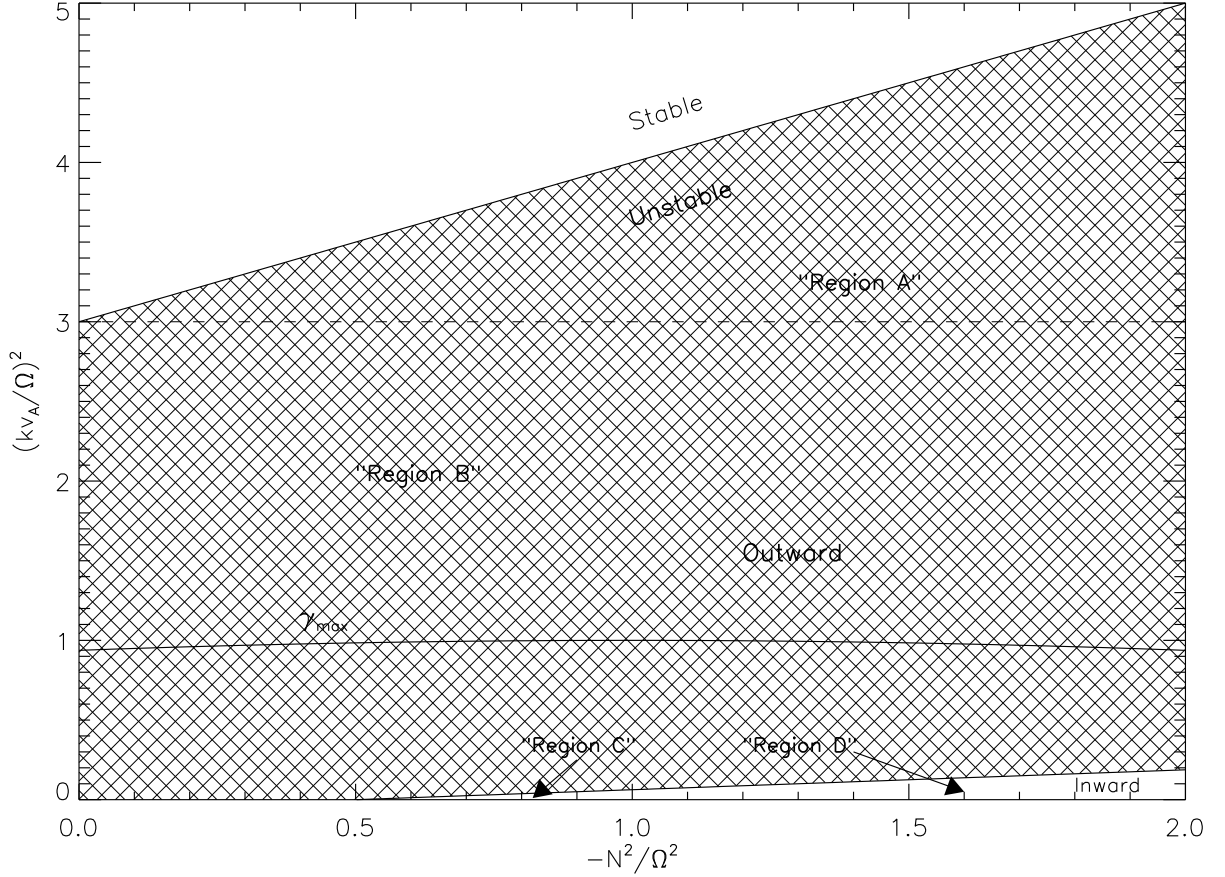}
\caption{ Region of instability in the $(kv_A/\Omega)^2$ --
$|N^2|/\Omega^2$ plane, for a magnetized disk obeying a Keplerian
rotation law.  (Other rotation profiles lead to qualitatively similar
diagrams.) The hatched region indicates the domain of outward
transport.  The solid line labeled $\gamma_{max}$ shows the wavenumber
of maximum growth rate as a function of ${\cal N}^2$.  The thin wedge
at the bottom corresponds to the inward transport region.  See the text
for the definitions of regions A through D.  } \label{mstability}
\end{figure}


Matters change radically when a magnetic field is present.  
By combining equations (\ref{growth}), (\ref{Rey}), and (\ref{Max}), we
may calculate the full range of wavenumbers that transport angular
momentum outwards:
\beq\label{out}
(kv_A)^2 > {1\over 16}\left(4-\left|d\ln\Omega^2\over d\ln R \right|\right)
\left(2{\cal N}^2 - \kappa^2\right).
\eeq
It is apparent that unless ${\cal N}$ is sustained at values in excess
of $\kappa/\sqrt{2}$, every mode, convective or otherwise, will
transport angular momentum outward.

The content of equation (\ref{out}) is shown graphically in figure (2).
Clearly, the dominant direction of angular momentum transport is now
outwards.  Note that the sense of transport is outward even for those
wavenumbers that satisfy the condition
\beq\label{conmd}
\left|{d\Omega^2\over d\ln R}\right| < 
(kv_A)^2 < {\cal N}^2 + \left| {d\Omega^2\over
d\ln R}\right|
\eeq
which are unstable {\em only} because of the presence of an adverse
entropy gradient.  These wavelengths occupy Region A in the figure,
specifically the shaded wedge above the dotted line $(kv_A)^2 =
3\Omega^2$.  These wavelengths would otherwise be stable to the ``pure
MRI.'' They are destabilized only because of the existence of adverse
entropy gradients, yet all region A modes transport angular momentum
outwards.  Physically this is because the large magnetic tension, which
regulates angular momentum transport in this regime, would oridinarily
be a strongly stabilizing agent.   The adverse entropy gradient
destabilizes, but does not alter the outward flow of angular momentum.

Region B denotes the shaded area below region A, but above
the unshaded wedge comprising regions C and D.  The B region is the
wavenumber domain of outward transporting modes that would be unstable
in the presence of the pure MRI.  Their growth rate is increased by the
presence of finite ${\cal N}^2$.  The curve showing the most rapidly
growing wavenumber (equation [\ref{mg}]) lies in region B, as shown, but
eventually leaves this region at larger values of ${\cal N}^2$ (see
below).

The inward transporting modes are confined to the unshaded narrow wedge
at the bottom of the diagram, corresponding to long wavelengths.  These
modes have only a small magnetic tension force, $(kv_A)^2 \ll
\Omega^2$.  Region C consists of modes with ${\cal N}^2 < \kappa^2$.
Comparison with figure (1) shows that region C modes are destabilized
only by the MRI (i.e., they are stable in hydro disks), but are
nevertheless associated with inward transport.  Finally, region D
identifies the modes that would be unstable in a purely hydrodynamic
system that likewise transport angular momentum inward.  In this narrow
domain, both magnetic tension and rotational stabilization are smaller
than the adverse entropy gradient.  The curve denoting the most rapidly
growing wavenumber enters region D for values of ${\cal N}^2$ in excess
of $4\Omega^2$.

The region of inward transport is very small, indeed non-existent for
${\cal N}^2<0.5 \kappa^2$, and it would be surprising if these
marginally-valid WKB modes effectively halted angular momentum
transport in nonradiative flows.  In the next section we show that this
is in fact impossible in any system where the seat of free energy is
differential rotation.  For the present, it is useful to have an
estimate (or a bound) of the size of ${\cal N}^2$ one expects in
nonradiative flows.

The entropy equation for a monotomic gas is
\beq
{3\over2} P {d\ \over dt} \left(\ln P\rho^{-5/3}\right) =  Q^{+}
\eeq
where $Q^+$ represents dissipative heating.  This will
generally not exceed
the total energy budget available from differential rotation,
$-T_{R\phi} d\Omega/d\ln R$, and may be much less.
In a one-dimensional approximation, we therefore expect
\beq
{3\over2} P v_r {d \over dr} \left(\ln P\rho^{-5/3} \right)  \lta
-T_{R\phi} {d\Omega\over d\ln R}
\eeq
Let us work near the equatorial plane and switch to cylindrical
radius $R$.
The inward drift velocity is related to the stress tensor by
(Balbus \& Hawley 1998):
\beq
v_R \simeq - {T_{R\phi}\over \rho R \Omega},
\eeq
which leads to
\beq
{1\over\rho} {d \over dR} \left(\ln P\rho^{-5/3} \right) \lta
{1\over3} {R\over P} {d\Omega^2 \over d \ln R},
\eeq
and
\beq
{\cal N}^2 = {3\over 5\rho} {dP\over dR} {d \over dR} \left(\ln
P\rho^{-5/3} \right) \lta {1\over 5} {d\ln P\over d \ln R}
{d\Omega^2\over dR}.
\eeq
Generally, when differential rotation increases, the pressure gradient
decreases, and vice-versa.  For a Keplerian profile, ${\cal N}^2/\Omega^2
\lta 0.6 d\ln P/d\ln R$.  In such a disk, pressure gradients are small,
and ${\cal N}^2/\Omega^2$ is likely to be less than unity.  If a
significant fraction of free energy goes into generating a magnetic
field that is carried off from the disk (i.e., not into dissipative
field reconnection), then ${\cal N}^2$ could be appreciably smaller.
The point here is that a well-defined bound on the rate of energy
dissipation limits the size of ${\cal N}^2$.

\section{Theoretical Implications} 

To the extent that an MHD fluid description is valid, the stability of
black hole accretion flow is regulated by the generalized H\o iland
criteria presented in Balbus (1995) or Balbus (2001).  The notion of a
``convective mode'' is ambiguous, as evidenced in figure (2).  A
perturbation in a given background flow is best characterized simply by
its wavenumber.  The important physical point is not to categorize and
study convective modes, but to understand that magnetic tension in an
accretion flow produces outward angular momentum transport.

\subsection {Turbulence and Irreversibility}

The dissipative properties of CDAFs are very striking, and, in light of
the results of \S2, they merit reexamination.  In a CDAF (NIA, AIQN),
outward angular momentum is driven by a viscous-like primary
instability in the fluid, and the energy from the differential rotation
is dissipated as heat.  This heating creates an adverse entropy
gradient which, in turn, generates a convective-like secondary
instability.  When both the primary viscous and secondary convective
instabilities are active, their angular momentum transport is equal and
opposite, and the total angular momentum flux drops to zero, as does
the volume-specific energy dissipation rate $Q^{+}$.  This means that
the convective instability recovers the dissipated heat (that
originally produced the convective instability), and returns it to the
fluid in the form of work (inward angular momentum and outward energy
fluxes).

Clearly, this is a violation of the second law of thermodynamics,
whether the system is hydrodynamical or magnetohydrodynamical:  the
onset of convection is caused by irreversible heat dissipation, and
this energy can not be fully recovered in the form of work.  
NIA identify 
\beq 
\label{T} - T_{R\phi} {d\Omega\over d\ln R} 
\eeq 
as the {\em fundamental} expression for the volume
specific dissipative energy loss rate.
But the fundamental
definition of the energy loss must be in terms of the dissipation
coefficients themselves.  With $\eta_v$ equal to the viscous
diffusivity, $\eta_B$ the resistivity, and $\delta {\cal J}$ the
fluctuating current density, the energy dissipation rate per unit
volume is
\beq 
Q^{+} \equiv \sum_i \langle \eta_v |\nabla \delta v_i|^2 +
\eta_B  {\cal \delta J}^2\rangle,
\eeq
where the sum is over vector components.  These losses never
vanish in a turbulent cascade, and a secondary source of turbulent
fluctuations can never fully recover the energy that is dissipated in
the process of triggering the creation of the same source.  It amounts
to reversing diffusion, a thermodynamic impossibility.

This effect is not small.  In a realizable nonradiative flow, none of
the entropy generated via dissipation ever leaves the system (except in
an outflow), for entropy can only increase.  The CDAF description
ignores all of the entropy production in the flow, entropy which would
be required to produce the vigorous convection driving angular momentum
inward.  Recall that ${\cal N}$ needs to exceed $\kappa/\sqrt{2}$ for
there to be any unstable modes transporting angular momentum inward,
and even this minimum threshold already requires substantial levels of
energy dissipation to be present.  The description of a CDAF as a
dissipation-free flow (or even nearly so) lacks a fundamental
self-consistency.

The sources and sinks for turbulent fluctuations may be read off
from Balbus \& Hawley (1998), eq.~(89):
\beq
- T_{R\phi} {d\Omega\over d\ln R} + P\del\bcdot\bb{\delta v} - 
Q^{+}.
\eeq
The expression (\ref{T}) happens to be equal to dissipative energy
losses $Q^{+}$ under strictly defined conditions that may be inferred
from the above:  steady, local turbulence in which the work done by
pressure is negligible.  This is, in fact, a good description of the
behavior of turbulence in magnetized differentially rotating flows.
An immediate consequence is that the time-averaged value of $T_{R\phi}$
must be $ > 0$, i.e.,
\beq
T_{R\phi} = - \left(d\Omega\over d\ln R\right)^{-1} Q^{+} > 0.
\eeq
The unavoidable presence of dissipation compels a net outward angular
momentum flux in any time-steady magnetized system in which the free
energy source is differential rotation.

\section{Comparison With Simulations}

The point of contact of CDAFs with numerical simulations is the power
law scaling of the density, $\rho(r)$, where $r$ is spherical radius.
The argument runs that since the CDAF energy flux is conserved (no
dissipation), $\rho v^3 r^2$ is a constant.  With $v\sim r^{-1/2}$,
this gives $\rho \sim r^{-1/2}$ as well.  By way of contrast, constancy
of the mass flux $\sim \rho vr^2$ would give a much steeper $\rho\sim
r^{-3/2}$ power law, which is associated with an ADAF solution.  The
power law scaling $\rho\sim r^{-1/2}$ has been interpreted as evidence
in support of convection-dominated flows.

Since gas accreting into a black hole is almost certainly magnetized at
some level, there is very little to be gained by hydrodynamical
simulations:  the stability and transport properties of magnetized and
unmagnetized gases are simply too different from one another.  In
general, MHD simulations have not been supportive of CDAFs (Stone \&
Pringle 2000; Hawley, Balbus, \& Stone 2001; Hawley \& Balbus 2002).
The one exception cited is that of Machida, Matusmoto and Mineshige
(2001).  This simulation follows the evolution of a magnetized torus in
a Newtonian potential.  ``Convective motions'' are observed in the
subsequent accretion flow.  However, these authors draw no distinction
between what they refer to as convective motions and turbulence in
general.  Turbulence is, of course, inevitable in an MHD accretion
flow; it is hardly the defining signature of a CDAF.  The only
quantitative basis for the claimed agreement with CDAFs was the
observation that at one point in time the the density exhibited a
$r^{-1/2}$ power law behavior.  However, the flow was a highly
time-dependent, the $r$ dependency evolved, and a different power law
emerged at the end of the simulation.

Even if $\rho \sim r^{-1/2}$ scaling were unambiguously extracted from
a simulation, there is no compelling association with thermal
convection.  Under steady conditions, all nonradiative flows,
regardless of their level of dissipation, have a conserved mass flux
and a conserved total energy flux.  When the flow is turbulent, to
address the behavior of these fluxes one must speak in terms of
correlations of the flow quantities.   In the presence of fluctuations,
it is common to have weak correlations in the mass flux and much
stronger correlations in the energy flux; indeed, ordinary waves
display this sort of behavior.  The scaling $\rho \sim r^{-1/2}$ could
emerge, for example, if the correlation between pressure and velocities
in the energy flux were strong, and the magnitude of all velocity
fluctuations scales by the local virial velocity.  Neither the presence
or absence of thermal convection, nor the rate of energy dissipation is
relevant.

In any case, any similarity between theory and simulation in a radial
power law index is not an answer to fundamental dynamical and
thermodynamical inconsistencies.  Verification should run in the
opposite direction: in the course of producing MHD turbulence, any
numerical simulation should show on average a positive value for
$T_{R\phi}$, and significant levels of energy dissipation.

The notion of a convectively unstable mode is ill defined in magnetized
differentially rotating systems.  When the magnitude of the \BV
frequency ${\cal N}$ is less than $\kappa/\sqrt{2}$, all unstable modes
transport angular momentum outward.  If ${\cal N}$ is maintained above
this, equation (\ref{out}) and figure (2) show that a small range of
wavelengths on the largest scales transport angular momentum inwards.
(It is of course not at all surprising that sufficiently large values
of ${\cal N}$ can be chosen to render the dynamics of differential
rotation less important than convection.)  For ${\cal N} <\kappa$, the
inward transporting modes are destabilized by magnetic tension, for
larger ${\cal N}$ values the destabilization is by convection.  Small
wavenumber convective modes (Region A in figure [2]) always transport
angular momentum outwards, and overall transport remains overwhelmingly
outwards for values of ${\cal N} \sim \Omega$.

This is a very serious difficulty for CDAF models, which rely on a
significant level of inward angular momentum transport engendered by
convectively-driven modes.  Moreover, and most importantly, the claims
of vanishing stress and vanishing energy dissipation are inconsistent
with fundamental thermodynamic principles.  They are also inconsistent
with the need to maintain ${\cal N} > \kappa/\sqrt{2}$, a requirement
for the existence of any inward transporting unstable modes.  The
central feature of CDAF theory, a vanishing of the angular momentum
flux, is not possible if the seat of free energy is differential
rotation.  Characteristic power law scalings extracted from numerical
simulations are not evidence for low dissipation flow.

Black hole accretion features far more conventional and familiar fluid
dynamics.  Underluminous sources are likely to be a consequence of low
densities and lower than anticipated temperatures, as opposed to
dissipation-free turbulence.  MHD turbulence leads to outward angular
momentum transport and to a positive $R\phi$ stress component,
irreversible dissipative heating, mass accretion, and significant mass
outflow as well.  These are all manifest in numerical MHD
simulations.  The development of the spectral and energetic properties
will test these model flows more stringently.

\acknowledgements{M.~Begelman, O.~Blaes, and C.~Terquem provided
helpful comments on a preliminary draft of this paper, as did
R.~Narayan, though he is in disagreement with our conclusions.  This
research was supported by NSF grant AST-0070979, and NASA grants
NAG5-9266 and NAG-10655.}

\end{document}